\documentclass[%
 aps,
 prr,%
 amsmath,amssymb,
 reprint,
showkeys,
]{revtex4-2}

\usepackage[T1]{fontenc}
\usepackage[utf8]{inputenc}
\usepackage{amsmath}
\usepackage{amssymb}
\usepackage{graphicx}
\usepackage{refstyle}
\usepackage{multirow}
\usepackage[
  citecolor=blue,
  colorlinks,
  linkcolor=blue,
  urlcolor=blue,
]{hyperref}
\usepackage[dvipsnames]{xcolor}
\usepackage{dcolumn}
\usepackage{bm}
\usepackage{soul}

\begin{document}


\title{{Cooperators overcome migration dilemma through synchronization}}
\author{Shubhadeep Sadhukhan}
\email{deep@iitk.ac.in}
\affiliation{
	Department of Physics,
	Indian Institute of Technology Kanpur,
	Uttar Pradesh 208016, India
}

\author{Rohitashwa Chattopadhyay}
\email{crohit@iitk.ac.in}
\affiliation{
  Department of Physics,
  Indian Institute of Technology Kanpur,
  Uttar Pradesh 208016, India
}
\author{Sagar Chakraborty}
\email{sagarc@iitk.ac.in}
\affiliation{
  Department of Physics,
  Indian Institute of Technology Kanpur,
  Uttar Pradesh 208016, India
}

\date{\today}

\begin{abstract}
Synchronization, cooperation, and chaos are ubiquitous phenomena in nature. In a population composed of many distinct groups of individuals playing the prisoner's dilemma game, there exists a migration dilemma: No cooperator would migrate to a group playing the prisoner's dilemma game lest it should be exploited by a defector; but unless the migration takes place, there is no chance of the entire population's cooperator-fraction to increase. Employing a randomly rewired coupled map lattice of chaotic replicator maps, modelling  replication-selection evolutionary game dynamics, we demonstrate that the cooperators---evolving in synchrony---overcome the migration dilemma to proliferate across the population when altruism is mildly incentivized making few of the demes play the leader game.
\end{abstract}
\maketitle
\section{Introduction}
Cooperation has strong ethical, moral, philosophical, and even theological implications for the human~\cite{2013_N}. Its spatiotemporal evolution in a structural arrangement of finite number of agents playing strategic games~\cite{1992_Nowak_May_Nature} is an insightful phenomenon that exemplifies similar real-life phenomena in social~\cite{2006_Axelrod}, economic~\cite{2009_P_PTR, 2019_B_QJE}, physical~\cite{1991_SKY_AI}, and biological~\cite{2013_CG_TCB} systems. There are many different mechanisms~\cite{2006_Nowak, 2011_WCG} of imparting cooperation. In the network of agents, migration~\cite{2009_Helbing_Yu_PNAS,  2009_WNH_PNAS, 2010YWW, 2011Y, 2011_Roca_Helbing_PNAS, 2012_FN_JSP, 2012WFZW, 2012CSP, 2013BTA, 2018_LELG_NC} could be one such mechanism: While success-driven migration~\cite{2009_Helbing_Yu_PNAS, 2011Y}, aspiration-driven migration~\cite{2010YWW}, expectation-driven migration~\cite{2012WFZW}, risk-driven migration~\cite{2012CSP}, opportunistic migration~\cite{2013BTA}, and migration following a satisfying dynamic~\cite{2011_Roca_Helbing_PNAS} lead to cooperation; {\color{black}random or diffusive migration of the agents is expected to suppress cooperation by facilitating invasion by defectors~\cite{1991_DW_AN,1993_EL_AB}.}

Consider the following scenario: A population is divided into a finite number of groups or demes wherein a very large number of individuals---cooperator and defectors---are interacting with each other to play the prisoner's dilemma (PD) game~\cite{1965_RC}. Furthermore, let there be migration of individuals from one deme to the other. Effectively, what we have is a network of demes. One expects that any structured network of such demes, where the network structure represents the migration from one deme to the other, should have the population state with only the defectors as an evolutionarily stable state which is resilient against invasion by the cooperators. Now suppose that all the demes with the PD have only defectors left over time. With a view to establishing cooperation~\cite{2003_FF_N}, in some of the randomly selected demes one encourages altruism by rewarding additional payoff to the cooperators who play against the defectors such that the PD transforms into the leader game (LG)~\cite{1967_R_BS} in the selected demes. The LG can be thought of as the modification of the PD such that an altruistic behaviour is rewarded, {i.e.}, a cooperator playing against a defector is given some additional payoff so that the resultant payoff is greater than the payoff for mutual cooperation. Note that a similar idea~\cite{2003_Skyrms} of punishing players who defect against the cooperators transforms the PD to the stag-hunt game. With {random} migration in action, can the LG induce sustained cooperation in the PD at the other demes?

An interesting dilemma arises: In the population with some demes having both the cooperators and the defectors playing the LG and some having exclusively defectors playing the PD, the cooperators would not want to migrate to the demes playing the PD lest they should be exploited. However, if they stay at the demes where the agents play the LG, the fraction of the cooperators can not increase throughout the population and they would be surrounded by a lot of defectors present in the other demes. This means that the cooperators would always be at the risk of being exploited by a free rider. To refer to this situation, we introduce the phrase, \emph{migration dilemma}, which is fundamentally different from other known social dilemmas like the tragedy of commons (TOC)~\cite{Hardin1243} and agglomeration dilemma~\cite{2011_Roca_Helbing_PNAS}. In view of this dilemma, it is not obvious \emph{a priori} whether the random migration helps in increasing the cooperator-fraction of the entire population.

The stylized game of the PD presents probably the simplest possible abstraction and visualization of the problem of emergence of cooperation: The players of the PD defect to play the non-Pareto-optimal Nash equilibrium~\cite{1896_Pareto,1950_N_PNAS} even though mutual cooperation would have fetched more payoff. Thanks to the folk theorem~\cite{2003_C, 2014_CT_PNAS} of the evolutionary game theory, through similar games one can also see how the game theoretic equilibria (e.g., Nash equilibrium) and the equilibria (e.g., fixed point) of corresponding dynamical systems---especially the paradigmatic replicator equation~\cite{1978_TJ_MB, 1983_SS_JTB,1998_HS, 2002_PN_JTB, 2003_C, 2005TCH}---are related. The real world, however, is enormously more complex: The evolutionary dynamics have other outcomes like chaos~\cite{1992_S_PSA, 1993_NS_PNAS, 2002_Sato_Farmer_PNAS,2011_V_PRL} that appears in the continuous replicator equation with more than two strategies. Interestingly, chaotically changing payoffs~\cite{2006_Perc_EPL} may enhance cooperation; and in turbulent aqueous environment, chaotic flows induce migration that may facilitate evolution of colonies via cooperation~\cite{2020_KSN_NC}.

\section{The Model}
In order to allow for rich dynamical complexities in our study and with a view to establishing the intriguing interplay between the random migration and the chaos, we consider a spatiotemporal model where in each deme a discrete replicator dynamic (replicator map) is in action and the migration-induced coupling between the demes presents us with a coupled map lattice (CML) of the replicator maps. The CMLs have been extensively studied~\cite{1992_Kaneko_Chaos, 2014_Kaneko} in the context of biological and computational networks, fluid dynamics, ecology, chemical reactions, \emph{etc.} It however is the most known for the study of spatiotemporal chaos in spatially extended systems. It is also natural that depending on the coupling strength between the lattice sites of a CML, synchronized dynamics~\cite{2001_P, 2014_Kaneko} may appear so that all the lattice sites evolve in unison.

\begin{figure}
	\includegraphics[scale=0.06]{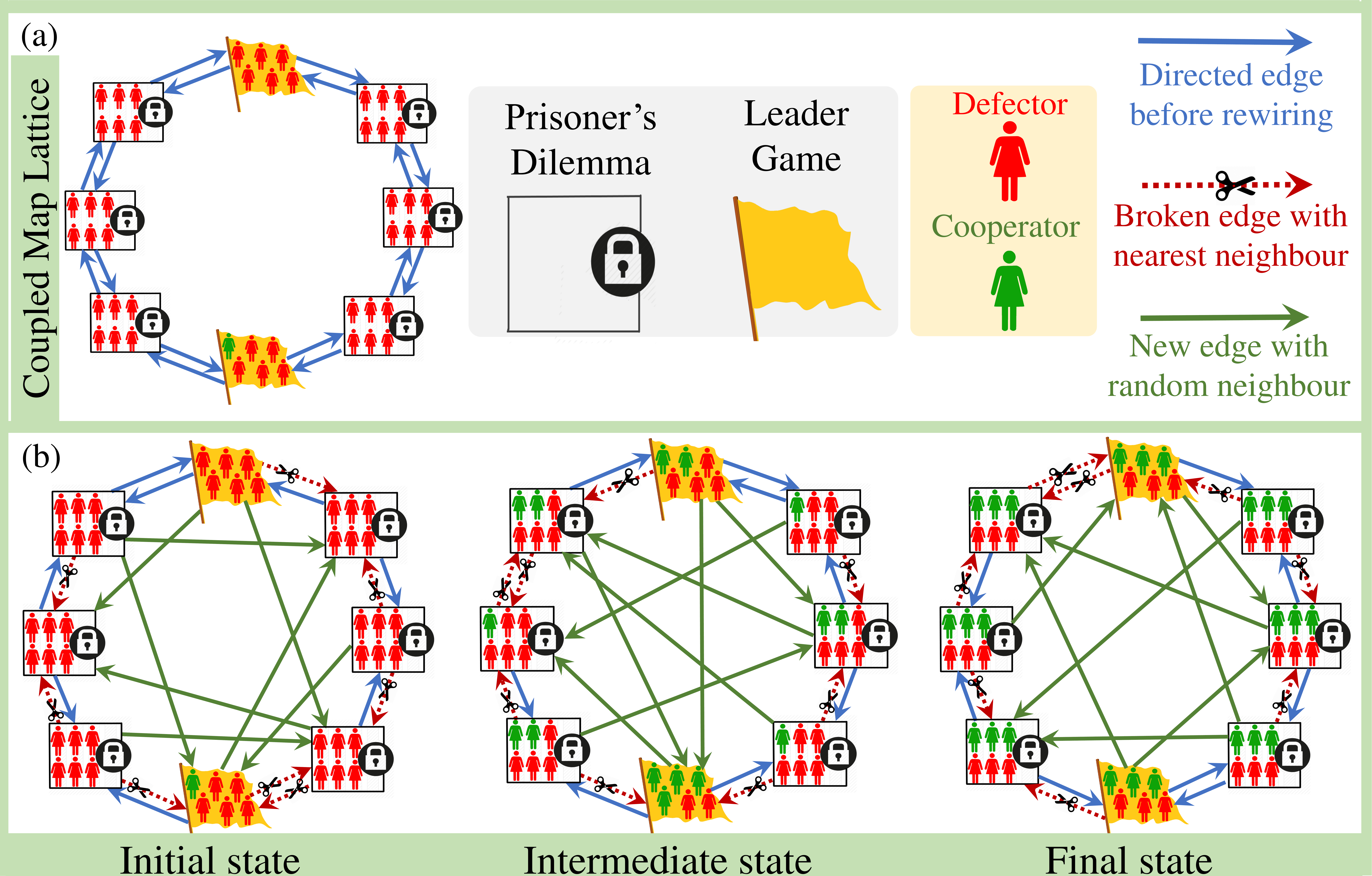}
	\caption{{Schematic diagram of CML with dynamic random rewiring.} We see in top row {(a)} the base CML with eight demes each having cooperators (green individuals) and/or defectors (red individuals). We exhibit only six representative individuals in each deme for illustrative purpose. Every deme has a game---say, the PD (square with lock) or the LG (orange flag)---played by the individuals in it. The arrowheads point towards the destinations of respective migration. In bottom row {(b)}, as dynamic random rewiring is employed, some of the directed edges (shown by blue arrows) of the base CML are randomly broken (as shown by red arrows with scissors) and new incoming edges (shown by green arrows) are created. The dynamic random rewiring is employed at each step of the time evolution---initial, intermediate, or final---helping in establishing enhanced cooperation throughout the CML with time, even after starting with a very small fraction of the cooperators at only one deme.}
	\label{fig:schematic_diagram}
\end{figure}

Thus, in the context of the migration dilemma, we construct a CML---as schematically presented in Fig.~\ref{fig:schematic_diagram}---where every lattice site is a deme in which the dynamics of the fraction of the cooperators is governed by the replicator map corresponding to either the PD or the LG. The replicator map corresponding to the LG has chaotic dynamics implying coexistence of the defectors and the cooperators. In the presence of the random migration, fashioned by temporal rewiring~\cite{2020_LZSCLWL_NC} of the edges of the CML, we wonder if the LG induce cooperation in the PD at other demes.

The one-dimensional replicator map~\cite{1997_BS_JET, 2000_HS_JEE, 2000_BV_QJE, 2003_C, 2010_M_AEJM, 2011_V_PRL,2018_Pandit_etal_Chaos, 2020_MC_JTB, Archan_2020_chaos},
\begin{equation}
	x_{n+1}=f(x_n):=x_n+x_n[({\sf A}\mathbf{x_n})_1-(\mathbf{x_n})^T{\sf A}\mathbf{x_n}], \label{eq:RM}
\end{equation}
such that $0\le x_n\le1$ for all $n$, for the two-player-two-strategy games is the simplest and most convenient testbed of our idea because it models the Darwinian selection, its fixed points correspond to Nash equilibria and evolutionarily stable states through the folk and related theorems, and it is endowed with chaotic attractors. Here,  subscript `1' denotes the first component of vector ${\sf A}\mathbf{x_n}$, $n$ denotes the time step, where 
${\sf A}=
\tiny{\begin{bmatrix} 
1 & S \\
T & 0\\
\end{bmatrix}}
$ 
is the payoff matrix for a player in the two-player-two-strategy symmetric game. $S$ and $T$ are real numbers. ${\bf x}=(x,1-x)$ is the state of the population such that $x$ is the fraction of the cooperators and $1-x$ is the fraction of the defectors. 

We consider a CML that is a one-dimensional linear network with $N$ nodes/lattice sites and periodic boundary condition such that each lattice site or deme is connected to its two nearest neighbours. In order to implement the random migration in the system under investigation, we modify the couplings in the CML. At every time step, any node can either allow migration from its two nearest neighbours or from two other demes picked uniformly randomly. The probability of remaining coupled to the nearest neighbours is $1-p$, where $p$ is called the dynamic rewiring probability; `dynamic' emphasizes that the rewiring is happening at every time step. Mathematically, the mean field equation for the CML with the random coupling is given by,
\begin{eqnarray}
&&x^i_{n+1}=[(1+S){x_n^i}+(1-2S-T){x_n^i}^2+(S+T-1){x_n^i}^3]\nonumber\\
&&\phantom{x^i_{n}}\times(1-\epsilon)+\frac{(1-p)\epsilon}{2}(x^{i-1}_n+x^{i+1}_n)+\frac{p\epsilon}{2}(x^{\xi}_n+x^{\eta}_n),
	\qquad\label{eq_rewired_network}
\end{eqnarray}
where, the superscript $i$ denotes the $i$th lattice site and $\epsilon$ is the coupling strength measuring the rate of migration to the $i$th node from the respective nodes. The coupling strength $\epsilon$ is equivalent to a diffusion coefficient {\color{black}(see Appendix~\ref{App:A})}. Furthermore, $\xi$ and $\eta$ are the indices of the two randomly chosen demes and are not equal to $i-1$, $i$, or $i+1$. We must restrict $\epsilon$ between $0$ to $1$ so that $x^i_n$ doesn't become either negative or greater than one and work with only those values~\cite{2018_Pandit_etal_Chaos} of $S$ and $T$ for which $0\le x^i_n\le 1$ for all values of $i$ and $n$. 

In our model with same ${\sf A}$ at all the nodes---all the demes having same game (say, the LG; see later)---there is a possibility that the dynamics at all the demes may be completely synchronized to an interior fixed point, \emph{i.e.}, $x^i=x^j=x^*$ for all $i$ and $j$. One could do a linear stability analysis~\cite{2002_Sinha_PRE,Rohit_2020_Chaos} to find whether this synchronized state is at all stable and hence attainable. As is shown in the next paragraph, such a stable state in fact exists when $\epsilon\ge\epsilon_{\rm crit}:=\left[(|df/dx|-1)/(|df/dx|-1+p)\right]_{x^*}$.  A convenient way of quantifying the extent to which the system is synchronized is to define, a global order parameter~\cite{2016_Wang}, $r_G:=|\sum_{i=1}^N e^{2\pi \sqrt{-1} x^i}|/N$, that should be unity asymptotically once the system attains complete synchrony. It is easy to note that for large $N$ and uniformly distributed $x^i$ in the interval $[0,1]$, $r_G=0$. Hence, any partially synchronized state have a non-zero value of $r_G$ that is less than unity. 

{\color{black}
To find  the critical coupling strength in the presence of dynamic random rewiring, we perform the linear stability analysis about a homogeneous fixed point by putting $x^i_{n}= x^* + \delta_n^i$ in Eq.~(\ref{eq_rewired_network}). Expanding the resulting expression up to first order to arrive at,
\begin{eqnarray}
\delta_{n+1}^i=\left(1-\epsilon\right)[(1+S)+2(1-2S-T)x^*\nonumber\\+3(S+T-1){x^*}^2]\delta_n^i+\frac{(1-p)\epsilon}{2}(\delta^{i+1}_n+\delta^{i-1}_n)\nonumber\\+\frac{p\epsilon}{2}(\delta_n^{\xi}+\delta_n^{\eta}).\label{1storder}
\end{eqnarray}
As an approximation, we consider the term consisting of random neighbours to be zero on average. This approximation is valid for the interior fixed points as the perturbations about them are equally likely to be either side of the fixed point, and hence average out to zero. (Although not necessary, smaller values of $p$ would further strengthen the approximation.) Consequently, being interested in the average synchronization level in an ensemble of different realizations of the CML, we henceforth drop the last term---${p\epsilon}(\delta_n^{\xi}+\delta_n^{\eta})/2$---of Eq.~(\ref{1storder}) from our calculations. Subsequently, writing the small perturbations as a sum of its Fourier components, $\delta_n^i=\sum_q \tilde{\delta}^q_n \exp\left(\sqrt{-1}qi\right)$, we arrive at the following expression:
\begin{eqnarray}
\label{phi_bound}
\frac{\tilde{\delta}^q_{n+1}}{\tilde{\delta}^q_n}=\left(1-\epsilon \right)[(1+S)+2(1-2S-T)x^*\nonumber\\+3(S+T-1){x^*}^2]+\epsilon\left(1-p\right)\cos q.
\end{eqnarray}
It is obvious that for the perturbations to die down, or in other words, for the fixed point to be stable, the modulus of the right hand side of Eq.~(\ref{phi_bound}) has to be less than 1. Keeping in mind that $0\le\epsilon,p\le1$, it means that if $|(1+S)+2(1-2S-T)x^*+3(S+T-1){x^*}^2|\leq 1$, then for every $\epsilon$ between $0$ to $1$, the fixed point is stable; but if $|(1+S)+2(1-2S-T)x^*+3(S+T-1){x^*}^2|>1$, then to ensure that the perturbations die down, we require
\begin{eqnarray}
\left(1-\epsilon \right)|(1+S)+2(1-2S-T)x^*\nonumber\\+3(S+T-1){x^*}^2|+\epsilon\left(1-p\right)<1.
\end{eqnarray}
A little rearrangement yields the critical coupling strength ($\epsilon= \epsilon_{\rm{crit}}$), beyond which the fixed point must be stable, as 

\begin{equation}
\label{eq_critical}
\epsilon_{\rm{crit}}=\frac{|\frac{df}{dx}(x^*)|-1}{|\frac{df}{dx}(x^*)|-1+p}, 
\end{equation}
where $\frac{df}{dx}(x^*)=(1+S)+2(1-2S-T)x^*+3(S+T-1){x^*}^2.$
}

\section{Main Results}
\begin{figure}
	\includegraphics[scale=.55]{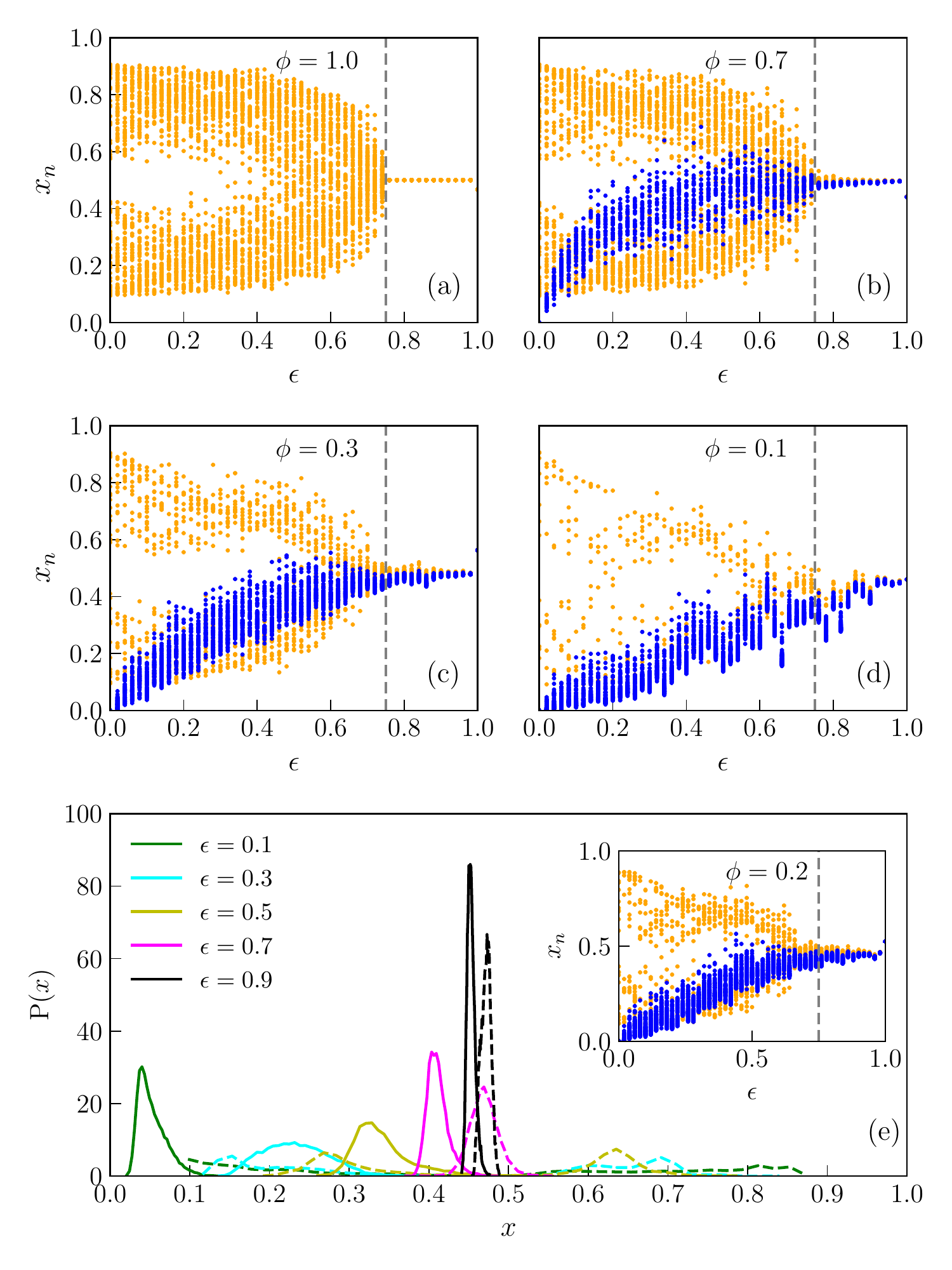}
	\caption{{Bifurcation diagrams: The demes with the LG induce cooperation in the demes with the PD.} We note in subplot {(a)} that if all the demes are playing the LG, after a critical value of coupling strength, $\epsilon=\epsilon_{\rm crit}$ (vertical dashed line) all the 100 chaotic trajectories (orange dots) synchronize onto the fixed point $x^*=0.5$ of the CML. As we introduce the PD in some of the demes with no (or some) cooperators, then the corresponding trajectories (blue dots) are pulled onto the synchronized state $x^*\approx0.5$ for all demes beyond $\epsilon_{\rm crit}$ as exhibited in subplots {(b)}-{(d)} for the LG game fraction, $\phi=0.7$, $0.3$, and $0.1$ respectively. In subplot {(e)}, having $\phi=0.2$, we note how the normalized the probability density function ($P(x)$) of the cooperator-fraction ($x$) for two randomly selected demes---one with the LG (solid lines) and the other the PD (dashed lines)---peak together near $x=0.5$ as $\epsilon$ increases.}
	\label{Fig:bifurcation}
\end{figure}

First consider that the demes of the CML are all of same type, \emph{i.e.}, same game is played at all the demes. For the purpose of this paper, the LG is of particular interest to us. Its payoff matrix is characterized by the inequality---$T>S>1>0$ (compare with the PD where $T>1>0>S$). In particular, we choose $T=8$ and $S=7$ as these values lead to the chaotic solutions~\citep{2018_Pandit_etal_Chaos} for the replicator map, given by Eq.~(\ref{eq:RM}), corresponding to the LG. Also, all the homogeneous fixed points, \emph{viz.}, $x^i=x^*=0$, $x^i=x^*=0.5$, and $x^i=x^*=1$ (for all $i$) are unstable under replicator map dynamics when dynamic rewiring is not employed.  As the strength of migration increases such that $\epsilon$ is more than $\epsilon_{\rm crit}$ (which is $0.75$ for $p=0.5$), the chaotic maps synchronize and all the demes have fifty percent cooperators in them (see Fig.~\ref{Fig:bifurcation}({a})).

In contrast with the above scenario, when the PD is being exclusively played at all the demes, we note that it has two only homogeneous fixed points, \emph{viz.}, $x^i=x^*=0$ and $x^i=x^*=1$ for all $i$. The former one is stable and the later one is unstable when the dynamic rewiring is not in action. Following a closer inspection of Eq.~(\ref{eq_rewired_network}), one may intuit dynamic rewiring modelling random migration to have two effects: reduced coupling strength ($\epsilon-p\epsilon$) with nearest neighbours and multiplicative noise of order $p\epsilon$. Therefore, almost any initial condition of the CML is attracted towards the all-defect state and as the corresponding phase trajectory approaches $x^i=0$ (for all $i$), the noise  becomes progressively weaker. Thus, the synchronized state of the CML is the state where none of the demes have even a single cooperator. If, however, in some demes the PDs transform to the LGs, we have the interesting situation where the CML has now mixed types of demes. With random migration in play, can demes with the LG induce cooperation in demes with the PD and hence help in overcoming the migration dilemma? 

In the CML with the two types of demes (\emph{i.e.}, ${\sf A}$, and hence $S$ and $T$ depend on $i$~\cite{2018_Hilbe_etal_2018, 2019_SMWN_PNAS}), let the fraction of the demes with the LG be denoted by $\phi$. It is quite evident that for any $\phi$, such that $0<\phi<1$, there should be competition between the demes playing the LG and the ones playing the PD in order to sustain cooperation. However, since the dynamics is chaotic, there always must be some cooperator at all times at all the demes even if to begin with the cooperators were present only in the demes with the LG. The dynamics would go towards a chaotic attractor that is being constantly perturbed due to the small effective noise generated by the dynamic random coupling. With the increase in the migration rate or the coupling strength, it may be noted that the cooperator-fraction increases in the demes playing the PD, and beyond the threshold, $\epsilon_{\rm crit}$,  it increases synchronously so much so that all the nodes of the CML have almost $50\%$ of cooperators.  This is shown in Fig.~\ref{Fig:bifurcation}; especially, Fig.~\ref{Fig:bifurcation}({e}) where one sees that the probability density functions of cooperator-fractions for two arbitrary demes---one playing the PD and the other the LG---almost merge as the coupling strength increases. What is remarkable is the fact that even with low values of $\phi$ (say, $\phi=0.1$; see Fig.~\ref{Fig:bifurcation}({d})) and with no initial cooperators in the demes playing the PD, random migration leads to strong emergence of cooperation in all the demes. Without any loss of generality, for the sake of concreteness, we have chosen $S=-0.1$ and $T=1.1$ for the PD.

 In short, beyond the critical migration strength, once the dynamics of demes are all almost synchronized and the average cooperation is about fifty percent, the cooperation is sustained at all times. The final value of $\langle \bar{x}\rangle\approx0.5$ is an enormous increase when compared with the initial value of $\langle\bar{x}\rangle\sim10^{-5}$ used in some of the runs of our simulations {\color{black}(see Appendix~\ref{App:B}).}
Here the angular brackets denoting average over many realizations of random migration and overbar denoting the average over demes. Thus, synchronization overcomes the migration dilemma. 
 
\begin{figure}
	\includegraphics[scale=.91]{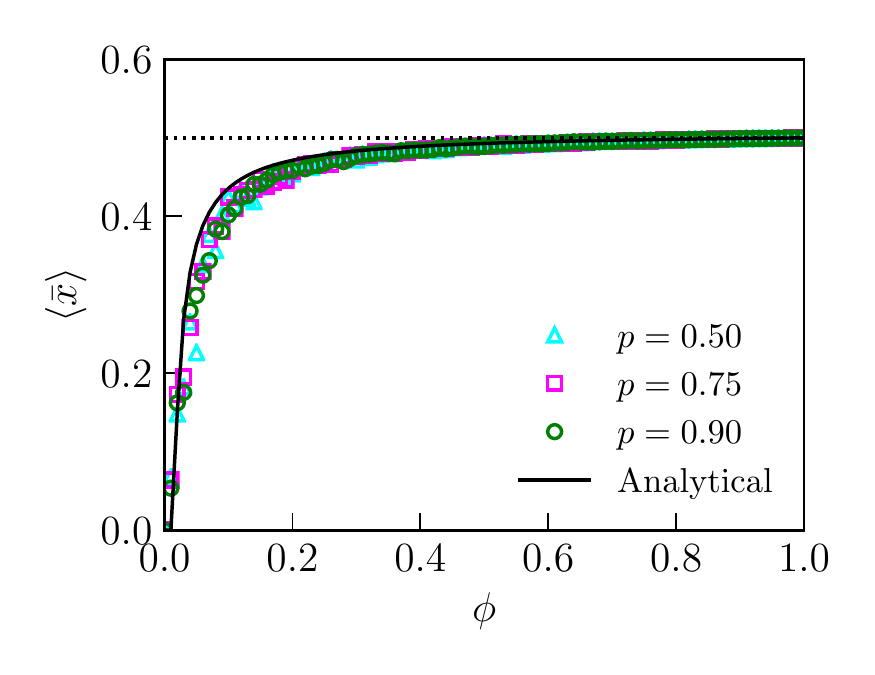}
	\caption{{Numerically validated analytical estimation of cooperation in the CML.} The cooperation level---characterized by $\langle\bar{x}\rangle$, the average cooperator fraction at each deme averaged over realization---in statistically steady state of the CML is plotted against the leader game fraction, $\phi$. Black solid line is the analytical estimation given by Eq.~(\ref{eq:AE}), whereas the markers denote the numerically calculated values; the cyan triangles, the magenta squares, and the green circles respectively indicate the dynamic rewiring probabilities, $p=0.5$, $0.75$, and $0.9$.}
	\label{Fig:theory_result}
\end{figure}

Since the level of cooperation can at most be fifty percent, the final synchronized state achieved in the system---apart from rendering much needed emergence of the cooperators---establishes the co-existence of strategies and hence, promotes biodiversity. In such states of the population modelled by the CML with the mixed types of demes, we can estimate the average cooperator-fraction. Since we are interested in a statistically steady homogenous state, suppose that the average state of each deme is by $\langle\bar{x}\rangle$. Furthermore, let $f_{\rm LG}$ and  $f_{\rm PD}$ denote the replicator maps respectively corresponding to the LG and the PD. Therefore, we expect that the effective replicator map for any deme should be a weighted average of these maps, and hence we expect
\begin{equation}
	\langle\bar{x}\rangle=\phi  f_{\rm LG}(\langle\bar{x}\rangle)+(1-\phi)f_{\rm PD}(\langle\bar{x}\rangle).\label{eq:AE}
\end{equation}
On solving this equation, we get a non-trivial solution for $\langle\bar{x}\rangle$ that is plotted in Fig.~\ref{Fig:theory_result} as a function of the fraction, $\phi$, of demes playing the LG. We note that it remarkably matches with the numerical simulations' results done at three different dynamic rewiring probabilities---$0.5$, $0.75$, and $0.9$. For the numerical simulations, we chose $\epsilon=0.9$ so that for all the three aforementioned values of $p$, synchronization is effected. It is interesting to note that the results are statistically independent of the exact value of $p$.

{\color{black}

\section{Synchronization suppresses migration dilemma: a robust mechanism}
The way cooperators overcome migration dilemma through synchronization is actually a very robust phenomenon. With a view to justifying this claim, we show in this section that how the phenomenon is independent of the other payoff matrices (that lead to chaotic dynamics) and also how the phenomenon shows up even if we use an evolutionary dynamic other than the replicator map.
\subsection{Other payoff matrices}
\begin{figure}[b]
\centering
\includegraphics[scale=0.85]{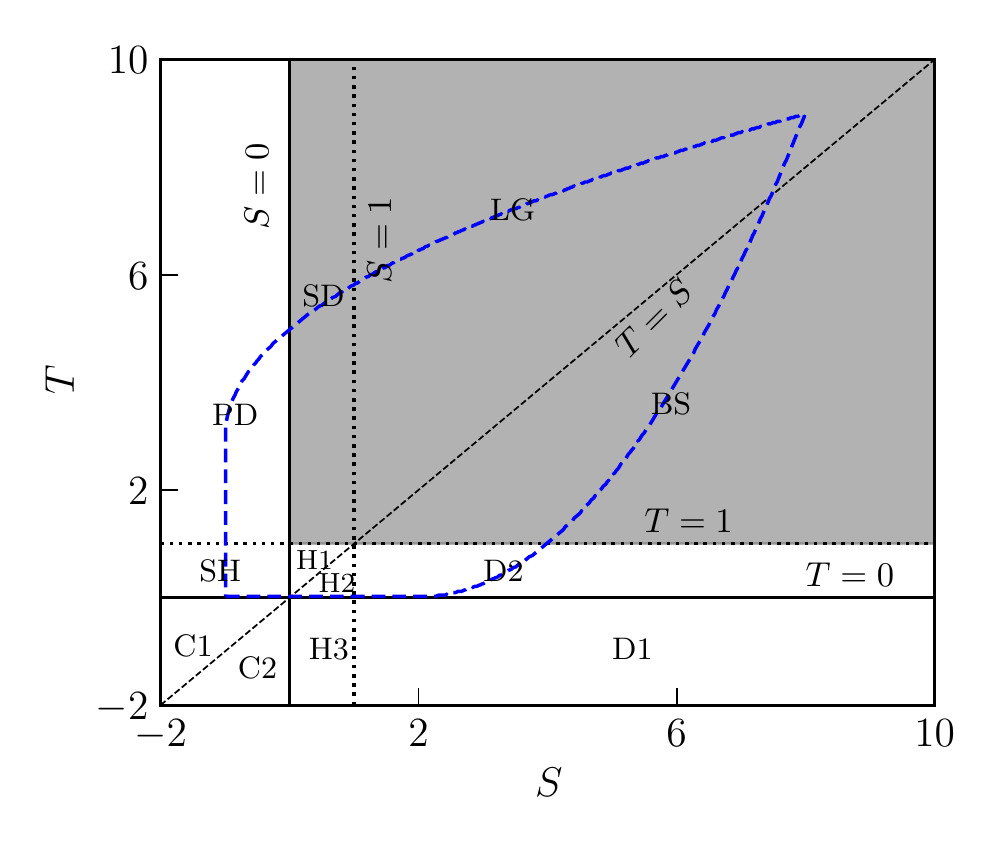}

\caption{Two-player--two-strategy symmetric games catagorised into twelve games in $S$-$T$ space. The grey shaded part of the picture denotes the anti-coordination games. Here PD: Prisoners Dilemma, SD: Snowdrift, LG: Leader game, BS: Battle of sex, SH: Stag Hunt, H1: Harmony I, H2: Harmony II, D2: Deadlock 2, C1: Coordination I, C2: Coordination II, D1: Deadlock I. The interior of the blue dashed curve denotes the allowed range of parameters that lead to physical solutions for the replicator map~\cite{2018_Pandit_etal_Chaos}.}
\label{fig_classification}
\end{figure}
Two-player--two-strategy symmetric games can be classified into twelve distinct games~\cite{2014_Hummert_etal_MBS} in terms of ordinality as shown in the Fig.~\ref{fig_classification}. We are mainly interested in the anti-coordination games (that consists of the leader game, the snowdrift/chicken game, and the battle of sexes) as  they are capable of giving rise to chaotic dynamics~\cite{2011_V_PRL, 2018_Pandit_etal_Chaos} when used in the replicator map. We recall that we have used the payoff matrix in the form ${\sf A}=
\tiny{\begin{bmatrix} 
1 & S \\
T & 0\\
\end{bmatrix}}
$. We used $S=-0.1$ and $T=1.1$ for the prisoner's dilemma (PD) game. We rewarded the altruistic behaviour in a fraction of the lattice sites such that they are effectively involved in a leader game. To check the robustness of our model against the nature of the game, we have simulated our model with the two other anti-coordination games---the chicken game and the battle of sexes.

 We have used $T=8$ and $S=7$ for the leader game for which the Lyapunov exponent, $\lambda=\frac{1}{n}\sum_{i=0}^{n-1}{\rm log} |f'(x)|$, is approximately 0.49 when used in replicator map. The positive Lyapunov exponent indicates chaos. For the {battle of sexes} game, we have used $T=6.45$ and $S=6.60$;  the corresponding $\lambda\approx 0.33$. In order to get chaos in the chicken game, we have taken a linear transformation of the payoff matrix---{\sf A} $\rightarrow$ $\gamma{\sf A}$, such that it remains a chicken game but gives rises to chaos. We have used $T=1.27$, $S=0.27$, and $\gamma=25$ for the chicken game~\cite{2018_Pandit_etal_Chaos} which then shows chaos as verified by a positive value of the Lyapunov exponent ($\lambda\approx 0.35$).

Now we have chaos for all these three games. In the Fig.~\ref{fig_trio}, we can see how the order parameter $\langle r_G \rangle$ and the average cooperation $\langle \bar{x}\rangle$ vary with the fraction $\phi$ corresponding to the anti-coordination game and the coupling strength $\epsilon$. In the Fig.~\ref{fig_trio}(a)--\ref{fig_trio}(c), it is shown how the system gets synchronized as the coupling strength is increased. The lower panel, Fig.~\ref{fig_trio}(d)--\ref{fig_trio}(f), the average cooperation is shown for three different games. We have calculated the critical coupling strengths for the synchronization calculated in the case of the CMLs with exclusively the anti-coordination games under investigation. The critical coupling strengths are $0.73$, $0.67$ and $0.75$ for the chicken game, the battle of sexes, and the leader game respectively. The interior fixed point corresponding to these there games are $0.50$, $0.55$ and $0.50$ respectively. So one can see that beyond the critical coupling strength, the average cooperation levels are fixed at the interior fixed point of the chaotic game when the fraction $\phi$ is not very low.

\begin{figure*}
\centering
\includegraphics[scale=0.9]{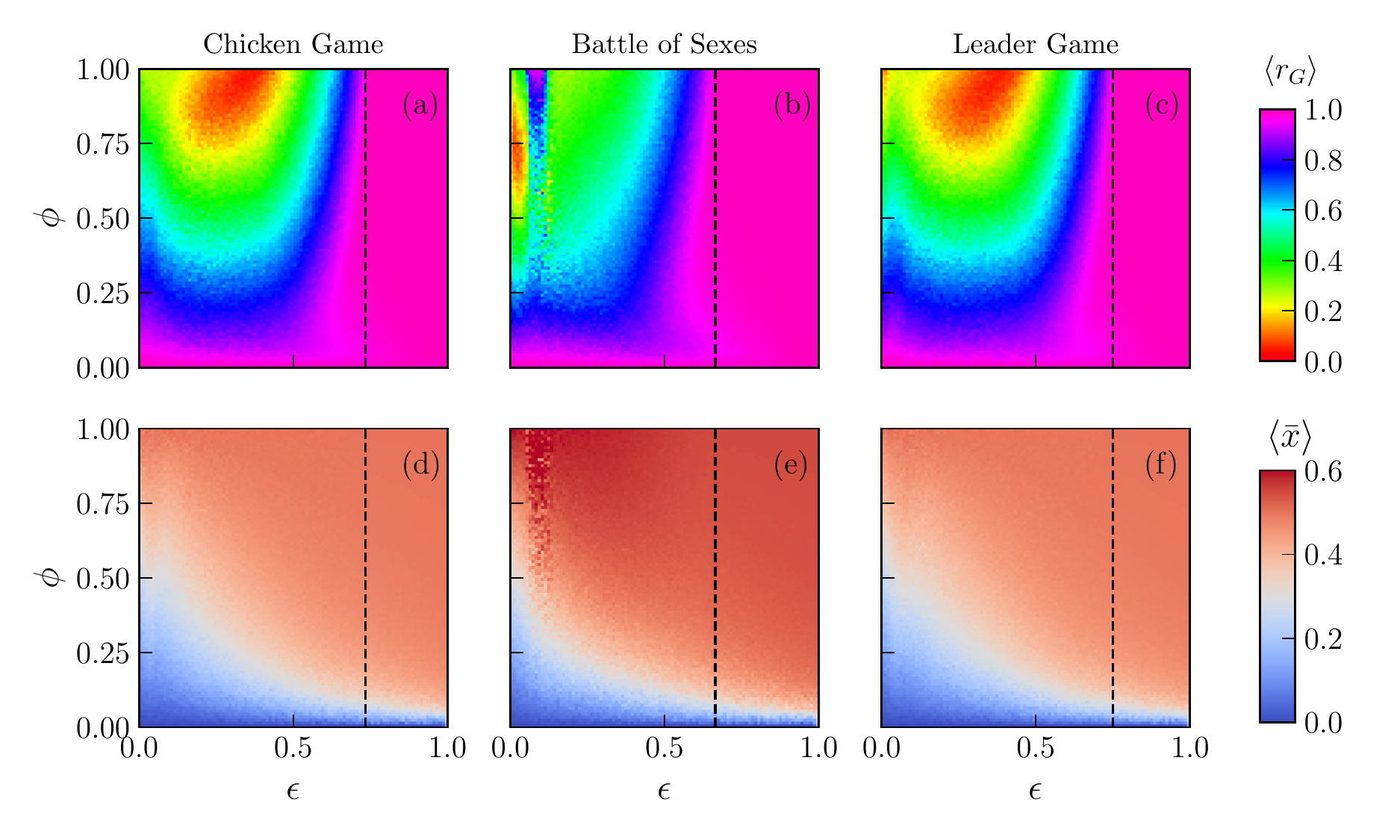}
\caption{In the upper panel (a)--(c), the order parameter $r_G$ is shown as a function of the coupling strength $\epsilon$ and the game fraction $\phi$ for the chicken game, the battle of sexes and the leader game respectively. We have shown the corresponding average cooperation over all the demes in the lower panel (d)--(f). Black dashed lines indicate the critical coupling strength when $\phi=1$. $p=0.5$ in these simulations.}
\label{fig_trio}
\end{figure*}

\subsection{Another evolutionary dynamics}

\begin{figure}
\includegraphics[scale=0.85]{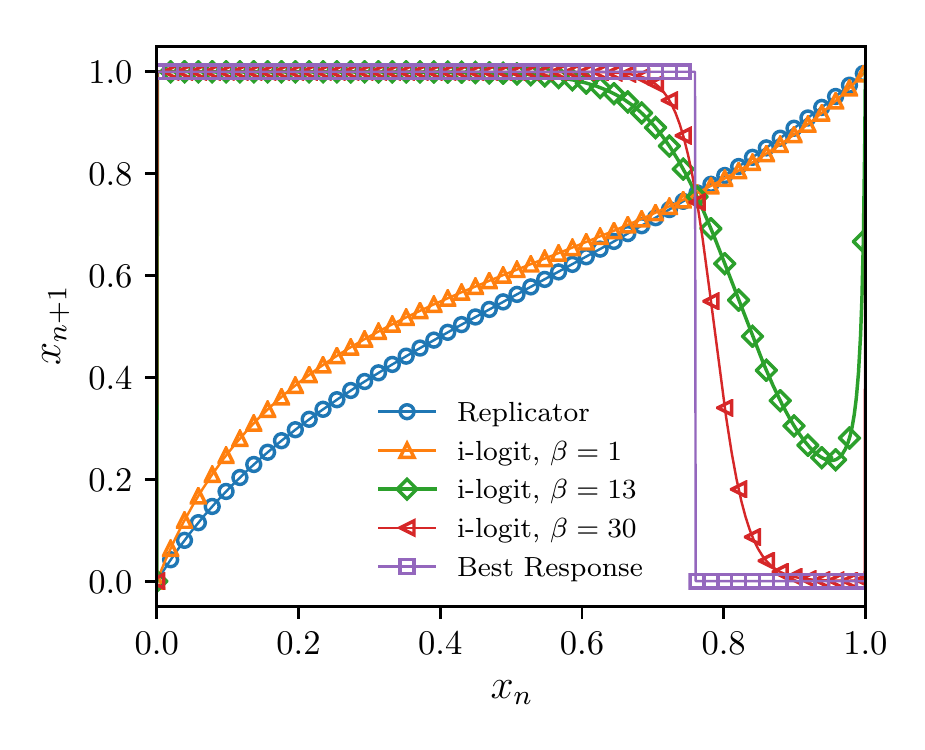}
\caption{Discrete time i-logit map's dependence on the myopic rationality parameter, $\beta$. $S=1.27$ and $T=1.4$ so that it corresponds to a leader game. At low $\beta$, resemblance with a version of replicator map, $x_{n+1}=x_n({\pi_C+k})/[{x_n \pi_C+(1-x_n)\pi_D+k}]$ (where $k$ is some background fitness in the absence of any competition); and at higher $\beta$, the resemblance with the best response dynamics, $x_{n+1}=\mathcal{H}(\pi_C-\pi_D)$ (where $\mathcal{H}$ is the Heaviside step function), are quite obvious.}
\label{fig_logit_model}

\end{figure}

\begin{figure*}[!t]
\includegraphics[scale=0.95]{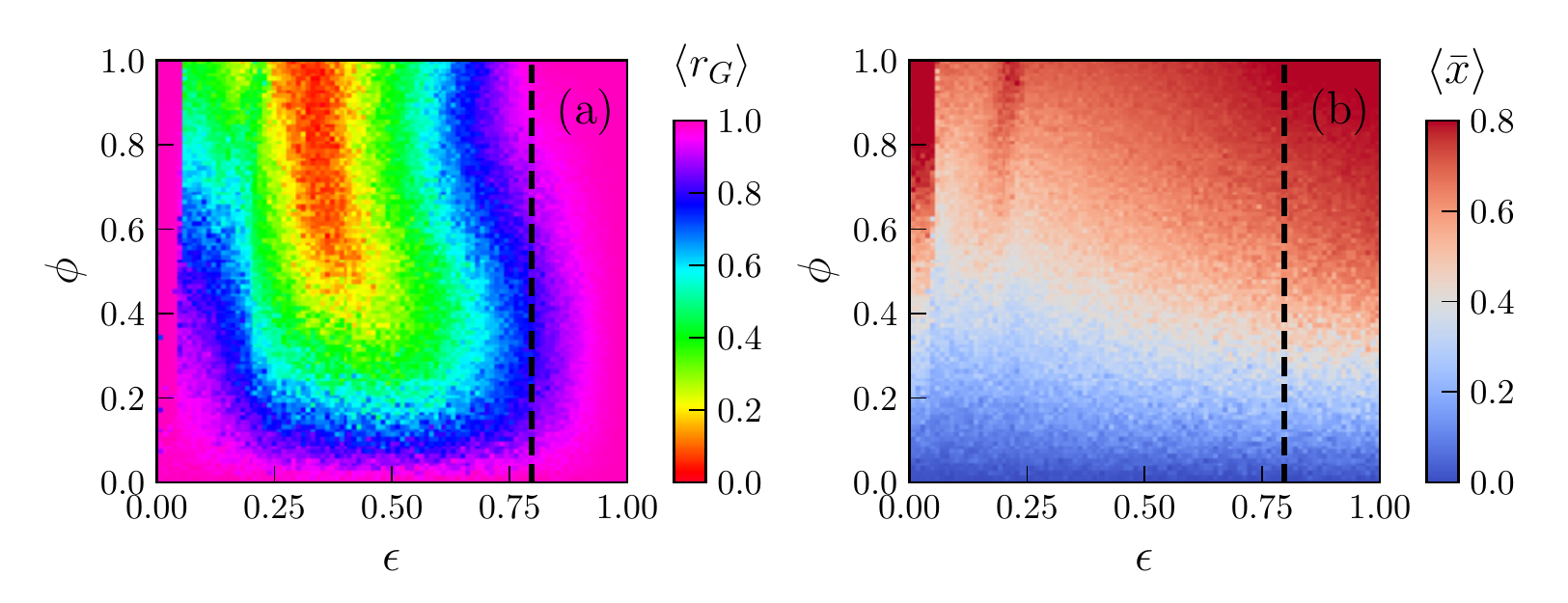}
\caption{(a) Order parameter and (b) average cooperation are shown as a function of the coupling strength $\epsilon$ and the chaotic game fraction $\phi$. Here we have used the parameters $S=1.27$ and $T=1.4$ for the leader game, the rationality factor $\beta=13$, and dynamic rewiring probability $p=0.5$.
}
\label{beta13}
\end{figure*}

The mechanism of synchronization overcoming migration dilemma is not restricted to the replicator dynamics. It is far more general. To establish this claim it suits us to present our studies with the imitative-logit (or i-logit) map~\cite{Cabrales_1992, Wagner2013}. 
For two strategy games, the discrete-time i-logit dynamic is given by the one-dimensional map,
\begin{equation}
\label{eq:i-logit_map}
x_{n+1}=g(x_n):=\frac{x_n}{x_n+(1-x_n)e^{\beta(\pi_D-\pi_C)}},
\end{equation}
where $\beta$, $\pi_C$, $\pi_D$ and $n$ are the rationality factor, expected payoff for player using first strategy (cooperating), expected payoff for the player using second strategy (defecting), and time step respectively. The expected payoffs for two strategies $\pi_C$ and $\pi_D$ are given by, $x_n+(1-x_n)S$ and $Tx_n$ respectively at the time step $n$.

Our main reason for choosing this model is the fact that i-logit map has a parameter, the myopic rationality factor $\beta$, that when varied can mimic various dynamics ranging from a version of the replicator map ($\beta\rightarrow0$) to the best-response dynamics ($\beta\rightarrow\infty$). This rationality factor models the ability of a player to choose strategy in such a way that maximizes her payoff. Higher $\beta$ implies lesser chance of mistake in choosing the wrong strategy. Thus, it is not surprising that high $\beta$ corresponds to the best response dynamics (myopically rational), while for low $\beta$ i-logit somewhat follows the replicator dynamics (low rationality) as shown in Fig.~\ref{fig_logit_model}.

Now we use this dynamics at the each site of the CML. We have mean field equation for the $i^{th}$ site is given by,
\begin{equation}
\label{eq:?}
x_{n+1}=(1-\epsilon)g(x_n^i)+\frac{\epsilon(1-p)}{2}(x_n^{i+1}+x_n^{i-1})+\frac{\epsilon p}{2}(x_n^{\xi}+x_n^{\eta}).
\end{equation}
where, $p$ and $\epsilon$ are the rewiring probability and the coupling strength respectively. $\xi$ and $\eta$ are the randomly chosen node index except $i-1, ~i$, and $i+1$. Using this i-logit model we get similar outcome as the replicator dynamics. We use the same parameters for the prisoner's dilemma as used in the replicator map. We find a set of parameters for the leader game which put the dynamics of the i-logit map in the chaotic regime: Consequently, we set $T=1.4$, $S=1.27$, and $\beta=13$ so that it shows chaos in the absence of any coupling; the Lyapunov exponent is found to be $0.37$ (approximately).

We find the order parameter and the average cooperation with the varying coupling strength $\epsilon$ and the fraction of the leader game $\phi$ and exhibit them in Fig~\ref{beta13}(a) and Fig~\ref{beta13}(b) respectively. Here again we can see  that beyond a critical value of coupling strength we have synchronization in cooperator fractions at every deme. (The critical coupling strength for $\phi=1$ which is $\epsilon_{\rm crit}\approx0.797$.) As $\phi$ increases towards unity the synchronised value is tending to the interior fixed point of the leader game which is $\approx 0.76$ in this case. 

}

{\color{black}

\section{Migration dilemma and its suppression by synchronization: A closer look}
The system we have studied in this work may remind one of the interdemic models specified by a migration pattern (island model~\cite{1943_Wright_G}, stepping stone model~\cite{1964_Kimura_Wiess_Genetics}, or a mixture of the two) and a competition model (differential proliferation~\cite{1982_CA_PNAS}, differential extinction~\cite{1982_K_E}, etc.) in the set of demes. However, in the conventional studies, a deme has finite number of individuals and hence genetic drift~\cite{2009_Charlesworth} plays an important role. By choosing to work with the deterministic dynamical equation, we are effectively working with the set of demes each having practically infinite number of individuals; hence, the genetic drift is out of the consideration. Thus, we are able to exclusively focus on the intriguing interplay between the chaotic dynamics at the demes and the migration between the demes.

Moreover, while the cooperators and the defectors can have different rates of migration~\cite{2009_WNH_PNAS}, we have kept the rates same in this paper. This simplifying assumption not only reduces the number of parameters in the problem, but also has the advantage that it neutralizes the effect of a possible mechanism of bringing forth cooperation in which the cooperators outrun the defectors~\cite{2013_DKCDG_PNAS}. Moreover, we have implemented symmetric migration, i.e., same migration rates between different allowed pairs of demes; the asymmetric migration is mostly known to alter the stability and the resilience of the population state~\cite{2018_LELG_NC}. While models with the migration of defectors exploiting a population of cooperators have been found to suppress cooperation in the overall population~\cite{1991_DW_AN,1993_EL_AB}, a model with success-driven migration of individuals have shown to enhance cooperation in the overall population~\cite{2009_Helbing_Yu_PNAS}.  In such models, either migration of only defectors occurs (in the former), or individuals' migration is voluntary (in the latter). Our model employs random migration of individuals---both cooperators and defectors---between any two arbitrarily chosen demes, which leads to synchronization between all the demes in the network, thereby enhancing overall cooperation.
\subsection{Importance of random migration}

	\begin{figure}[!t]
		\includegraphics[scale=0.6]{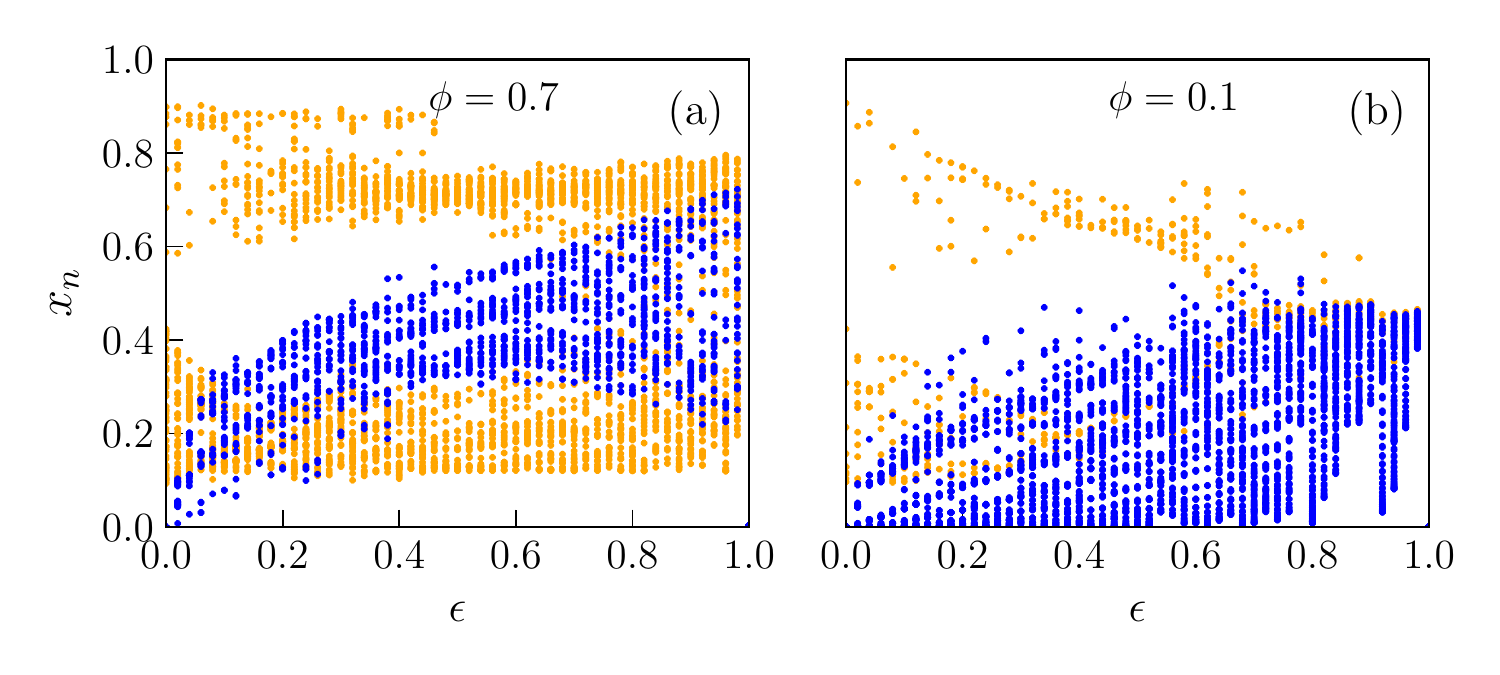}
		\caption{Bifurcation diagrams: Absence of synchronized cooperation in the CML with 100 lattice sites. Whether the fraction of LG is high (a) $\phi=0.7$ or low (b) $\phi=0.1$, the chaotic evolution of the cooperator fraction at the demes with the LG (orange dots) is not settling down into a stable synchronizied state with demes where the PD (blue dots) is being played. {The rewiring probability is fixed to $p=0.5$.}}
		\label{fig_without_random}
	\end{figure}

The incorporation of the random migration is very crucial for the results obtained in our paper. In order to explicitly show it, we present Fig.~\ref{fig_without_random} where we have simulated the evolution of the CML with demes playing either the LG or the PD but with the nearest neighbour migration only (i.e., no random migration); mathematically, the evolution is governed by Eq.~(\ref{eq_rewired_network}) with $p$ put to zero.

Fig.~\ref{fig_without_random} shows the cooperator fraction as a function of the coupling strength $\epsilon$ in all the demes of our network at a particular time step for given fractions ($\phi$) of the demes with the LG. We see that for both the low ($\phi=0.1$) and the high ($\phi=0.7$) values of the fractions, a majority of demes have fraction of cooperators way less than 0.5 for the entire range of the coupling strength. In contrast when random migration is present  (refer Fig.~\ref{Fig:bifurcation}) the cooperation fraction of all the demes converges towards 0.5 as $\epsilon$ is increased. This happens because, with random migration between the demes, the fixed point $x^*=0.5$, becomes stable while the other two fixed points---0 and 1---remain unstable. Thus, the random migration drives up the cooperator fraction from a low initial value to approximately 0.5 in the deme playing the LG, which in turn pulls up the cooperator fraction in the demes playing the PD to approximately 0.5. Interestingly, this synchronization to a fixed point is very robust, which does not get affected by all the defectors coming in from the neighbouring demes playing the PD. The migration of cooperators from the deme with the LG drives up the entire system's overall cooperation fraction to a higher level, starting from a low initial value; the system needs to be beyond the critical coupling strength that leads to the synchronization induced cooperator fraction's sustenance in all the demes to a value of around 0.5 for all time steps. This synchronization mechanism is absent when the migration is not random (recall Fig.~\ref{fig_without_random}). 

At first glance, one may be tempted to conclude that in a CML with both the LG and the PD, cooperation should trivially be established based on the following argument: Given that both cooperation and defection can be found long-term in the LG, but only defection in the PD, the defectors migrating from a deme playing the PD into a deme playing the LG would not threaten cooperation in the latter; on the other hand, given that cooperation can be found transiently in the PD, a continual arrival of cooperators from the LG to allow them to be found long term on the PD sites. The immediately preceding paragraph hints that it is not so straightforward because the aforementioned argument does not take into account the facts that how crucial the random nature of migration and the synchronization beyond a critical coupling strength are in the establishment and sustenance of cooperation in the CML; any arbitrary kind of migration and unsynchronized dynamics cannot result in sustenance of non-negligible cooperator fraction across all the demes at all times. While the continuous arrival of cooperators from the demes playing the LG demes indeed allows them to be found long term on the PD sites, we must appreciate that this continuous generation of cooperators at the LG demes and their arrival of cooperators in the demes playing the PD---thereby leading to the sustenance of cooperation---is achieved via synchronization that is established only when random migration is in action.
\subsection{Migration dilemma}
In the population with some demes having both the cooperators and the defectors playing the LG and some having exclusively defectors playing the PD, the cooperators would not want to migrate to the demes playing the PD lest they should be exploited. However, if they stay at the demes where the agents play the LG, the fraction of the cooperators can not increase throughout the population and they would be surrounded by a lot of defectors present in the other demes. This means that the cooperators would always be at the risk of being exploited by a free rider. The term ``migration dilemma" has been introduced by us to refer to this situation. As we discuss below, this is a very distinct kind of dilemma that has not been investigated earlier. First we recall a few well-known dilemmas.

The tragedy of commons (TOC)~\cite{Hardin1243} and its two-player version, arguably, the prisoner's dilemma~\cite{1965_RC}, constitute the most well known dilemma. The dilemma arises when the players choose to either cooperate (to various degrees) or defect with a plan to exploit a finite public good. The players get the maximum combined payoff if they cooperatively and restrainedly utilize the good. But if a player exploits (i.e., she defects while others cooperate) the good without restrain, she receives the maximum individual payoff. Therefore, all players would defect to lead to destruction of the public good. Another dilemma of interest is the agglomeration dilemma~\cite{2011_Roca_Helbing_PNAS}. In the model that leads to the agglomeration dilemma~\cite{2011_Roca_Helbing_PNAS}, the individuals in a network play the public goods game (PGG)~{\color{black}\cite{2009_WNH_PNAS}}. Some lattice points in the network are vacant for individuals to migrate to if they are not satisfied with the payoff received at their current locations. Now the dilemma the individuals are facing is whether they should agglomerate into large groups or not. This is because for a single PGG, with every individual's contribution, the benefit obtained decreases with increased group size. The benefit of an individual in large groups become more dependent on others' contributions, which is a risky proposition. Despite this risk, the potential for higher benefits also increases in a large group as an individual can get involved in more PGGs with its neighbours, thereby constituting a dilemma. The migration dilemma is different from these dilemmas. 

The TOC anyway does not involve any migration to have any direct resemblance with the migration dilemma. However, the migration dilemma in intertwined with the TOC: In the CML if the migration dilemma is not averted, the demes playing the PD game will be exploited off its resources and so a partial TOC will be effected in the demes. Random migration, assisted by synchronized dynamics, develops groups of cooperators competing with the defectors at every deme, so that the TOC may be averted.
	
Although it involves {migration} of individuals in a network, the agglomeration dilemma is fundamentally different from the migration dilemma. Unlike the case with the latter, there is no usage of chaos or synchronization in the formulation or the resolution of the agglomeration dilemma. Migration is a voluntary decision taken by an individual in the setup constituting the agglomeration dilemma, and all individuals do not need to migrate to form large groups. Whereas, migration of individuals from all the demes is a built-in feature of the setup showcasing the migration dilemma. Another essential feature in our model is that the migration between the demes is random (not based on any decision-process). In the case of agglomeration dilemma, the individuals can only migrate to specific empty sites in a lattice that fall within a certain predefined range from the individual's current location.

Probably, the most appealing feature of the migration dilemma is that it brings together the ideas of synchronization, cooperation, and chaos in the context of evolutionary game dynamics.

}

\section{Conclusions}
Summarizing, we have provided a macroscopic description of the emergence of cooperation owing to the synchronization of chaos in a population split into a network of demes having random migration among them. Specifically, we find that, by inducing synchronization, the random migration can actually increase cooperation in such a structured population. Interestingly, our model is seen bringing forth a stable biodiversity in the form of a heterogenous population (mixed cooperator-defector state) from an initially homogeneous population (all defector state) where some defectors in a few select demes are incentivized to cooperate. To the best of our knowledge this is the first demonstration of such constructive outcomes of chaos and synchronization in the theory of evolutionary games. This general mechanism is quite robust against the change in the payoff matrix and the game dynamics as long as chaos is observed
and distinct from the other migration induced cooperation models~\cite{2009_WNH_PNAS,2009_Helbing_Yu_PNAS,2013_DKCDG_PNAS,2018_LELG_NC}.

Another interesting angle is that the interconnected demes exclusively with the PD-players are subject to the (in)famous social dilemma of the tragedy of commons (TOC)~\cite{Hardin1243}: Any common resource therein would be exploited without restrain. Migration, even when random, goes a long way to develop groups of cooperators, at every deme, competing with the defectors so that the TOC may be averted. All it requires is that in a tiny fraction of the demes, the cooperation is encouraged; and migration propagates the virtue across at every other deme. Thus, while overcoming the migration dilemma, the cooperators also simultaneously tackle the dilemma pertaining to the TOC. In the light of the recent studies~\cite{2016_WEPBR_PNAS, 2019_Lin_Weitz_PRL, 2020_TPA_NC} on the TOC using feedback included replicator equation with payoff matrices from two different games (like in this letter), we believe that chaotic synchronization's hitherto unexplored utility in the TOC could be a useful future avenue of research.
\\%
\acknowledgements
The authors are grateful to Srihari Keshavamurthy, Archan Mukhopadhyay, and Supratim Sengupta for helpful discussions and invaluable suggestions.

\appendix
{\color{black}
\section{A note on the coupling strength}
\label{App:A}
The coupled map lattice (CML) is a very useful Eulerian description of dynamical systems to model a variety of phenomena in nonlinear systems. A CML consists of a collection of maps (discrete-time dynamics) that interacts with each other by means of a network structure among them. This interaction was modelled by simple diffusive coupling in the very early of its history~\cite{kaneko1984,waller1984}. As the simplest nontrivial case, consider that a particular map dynamics at a site of the lattice is given by $x_{n+1}=f(x_n)$. Then the equation for the $i^{th}$ site that is coupled to the two nearest neighbours through a diffusive coupling is given by,
\begin{equation}
x_{n+1}^i=f(x^i_n)+\frac{\epsilon}{2}(x^{i+1}_n-2x^i_n+x^{i-1}_n),
\label{eq_cml}
\end{equation}
where, $\epsilon$ is the coupling parameter. While $f(x)$ could be any map and $x$ could be any relevant physical quantity, for the specific case of the replicator map used in our paper, $x$ is the population fraction. We {immediately} note that {in} the continuum limit, we can change our site index $i$ to a continuous spatial variable $s$ and the time index $n$ to a continuous time variable $t$ to get the continuous equation,
\begin{equation}
\label{eq:continumm}
\frac{\partial x(s,t)}{\partial t}=f(x(s,t))-x(s,t)+\frac{\epsilon}{2}\frac{\partial^2 x(s,t)}{\partial s^2}.
\end{equation}
 From Eq.~(\ref{eq:continumm}), we can conclude that $\epsilon$ is a diffusion coefficient which can, in general, take any value between $0$ to $\infty$. Physically, larger the $\epsilon$ (i.e., diffusion coefficient), faster the diffusion; mathematically, if one waits for infinite time, $x$ at a site will completely diffuse out to neighbouring sites irrespective of the value of the diffusion coefficient. 
 
Eq.~(\ref{eq_cml}) however has a mathematical problem when compared with its continuum version, Eq.~(\ref{eq:continumm}): Since the coupling part represents the average of the population flux to the site $i$ from the connecting sites, it is reasonable to restrict $\epsilon$ between zero to unity. Still one can have unbounded solutions~\cite{yamada1983} for reasonable values of $\epsilon$. Although this drawback of the modelling is ignored  often, but in doing so we lose touch with the physical reality. One way~\cite{yamada1983} to recover the physical reality is to separate the diffusive processes from the reproductive process ($f(x)$). First, the new population in each cell is taken to be $x'^i_n=f(x^i_n)$ and this is followed by the diffusion process between the sites, so that the at the next generation we get
\begin{eqnarray}
&&x_{n+1}^i=x'^i_{n}+\frac{\epsilon}{2}(x'^{i+1}_n-2x'^i_n+x'^{i-1}_n),\nonumber\\
\implies&&x_{n+1}^i=(1-\epsilon)f(x^i_n)+\frac{\epsilon}{2}[f(x_n^{i+1})+f(x^{i-1}_n)].\qquad
\label{eq_modified_cml}
 \end{eqnarray}
Another way of enforcing physical solution is to slightly modify Eq.~(\ref{eq_cml}) as follows:
\begin{eqnarray}
x_{n+1}^i=(1-\epsilon)f(x^i_n)+\frac{\epsilon}{2}(x_n^{i+1}+x^{i-1}_n).
\label{eq_modified_cml2}
 \end{eqnarray}
 We note that if $\epsilon$ lies between 0 to 1, the righthand side of Eq.~(\ref{eq_modified_cml2}) is a convex combination of $f(x^i_n)$ and $(x_n^{i+1}+x^{i-1}_n)/2$ both of which are constrained to be between 0 and 1; and hence, $x_{n+1}^i$ must also always remain between 0 to 1 to help keep the solution always physical. Thus, the necessity of keeping the coupling parameter between $0$ to $1$ is to enforce physical reality, although in the continuum limit it can take any real value.
 It is worth pointing out that in the limit of weak selection $f(x)\approx x$ and both forms,  Eq.~(\ref{eq_modified_cml}) and Eq.~(\ref{eq_modified_cml2}), are approximately the same.

\section{Details of the numerical simulations}
\label{App:B}
{This section} describes the numerical methods used in producing the results reported in this paper. We used a parallel in-house C++ code and several libraries like ``blitz++'' for array operations, ``YAML'' for inputs, and ``MPI'' for task parallelism. Each deme/node/lattice site in the CML was an object of our user-defined class.\\
\\
\textit{Setup and initial conditions:} The underlying game, e.g., the PD or the LG, within a node was specified during the initialization of the system. We marked two games the LG and the PD by two numbers---$1$ and $2$ respectively. For a given fraction $\phi$ of the LG, we called a random number, $r$, between $0$ to $1$ using a uniform random number generator---drand48(). If $r\leq \phi$, then we assigned the game type value of the node as $1$, i.e., the LG; otherwise, game type was assigned $2$, i.e., the PD. Subsequently, we set the values for the payoff matrix, which is the attribute of a node. This is how we distributed two game types among the nodes for a given $\phi$.
For initial cooperator fraction in different node was assigned randomly by choosing a random number between $0$ to $1$ using a uniform random number generator. To test the robustness of our results, we also used some special initial conditions, e.g., zero cooperator-fraction at nodes with the PD and $x=0.001$ at all other nodes.
\\
\\
\textit{Dynamic random rewiring:} The CML used in this work is dynamic, i.e., it is a network such that its nodes' in- and out-degrees are stochastically changing over time. We began with a simple linear chain network where every node was connected to its two nearest neighbours with periodic boundary condition---in- and out-degree of each node was two---effectively creating a simple ring network. The incoming edges correspond to the immigration from the neighbouring nodes. At the beginning of every time step, we did the following rewiring of the simple ring network: For every node $i$, we generated a uniform random number $r^i$ between $0$ to $1$. If $r^i\leq p$, then the corresponding $i$th node was selected for rewiring; otherwise, the node remained connected to its nearest neighbours. If a node was selected for rewiring, then its incoming edges from its two nearest neighbours were deleted and new incoming edges were made with two randomly chosen nodes other than itself and its nearest neighbours. We repeated this process at every time step starting from the simple ring network. 
\\
\\
\textit{Probability distribution function calculation:} In order to compute the probability distribution as presented in Fig.~\ref{Fig:bifurcation}{(e)}, we saved the cooperator-fraction, $x^i$, at each node for $12000$ time steps. Neglecting first $2000$ transient time steps, we computed the normalized probability distribution using the data corresponding to the last $10000$ time steps using \emph{Python}. For presentation purpose, we chose two nodes randomly, one playing the PD and the other playing the LG. The probability distribution function P($x$), where P($x$)$dx$ is the probability of having the cooperator-fraction between  $x$ to $x+dx$ for each node under study. 
\\
\\
\textit{Parameters in numerical simulations:}
We used the CML with $100$ demes for every simulation and also checked for the robustness of the results with change in the CML's size. We varied the game fraction $\phi$ and the coupling strength $\epsilon$ from $0$ to $1$ in steps of $0.01$. For the PD, the variable payoff matrix elements were fixed at $T=1.1$ and $S=-0.1$, and for the LG we chose $T=8$ and $S=7$. The later shows chaotic dynamics in isolation. To verify that, we computed the maximum Lyapunov exponent for the replicator dynamics with the payoff matrix of the LG and got a positive value as expected. We simulated the system for $2000$ time steps in order to get a statistically steady state. Furthermore, as far as the cooperator-fraction and synchronization parameters are concerned, we did an average over $64$ different realizations in parallel using the MPI.

}

\bibliography{game_theory_shubhadeep.bib}
\end{document}